\documentclass[journal]{IEEEtran}
\usepackage[utf8]{inputenc}
\usepackage[T1]{fontenc}
\usepackage[cmex10]{amsmath}
\usepackage{booktabs}
\usepackage{multirow} 
\usepackage{xcolor,soul,framed} %,caption
\usepackage{url}
\usepackage[hidelinks]{hyperref}

\colorlet{shadecolor}{yellow}
\usepackage[pdftex]{graphicx}
\graphicspath{{../pdf/}{../jpeg/}}
\DeclareGraphicsExtensions{.pdf,.jpeg,.png}

\usepackage[cmex10]{amsmath}

\usepackage{array}
\usepackage{mdwmath}
\usepackage{mdwtab}
\usepackage{eqparbox}
\usepackage{amsmath}

\usepackage{amssymb}
\allowdisplaybreaks
\begin{document}
\bstctlcite{IEEEexample:BSTcontrol}
    \title{AFDM for LEO Inter-Satellite Links: Path-Level CSI Prediction and CRLB-Guided Pre-Equalization}
  \author{Houtianfu Wang,~\IEEEmembership{Student Member,~IEEE,}
      Ozgur B. Akan,~\IEEEmembership{Fellow,~IEEE}\\
      % <-this % stops a space

 \thanks{The authors are with Internet of Everything Group, Department of Engineering, University of Cambridge, CB3 0FA Cambridge, UK.}
  \thanks{Ozgur B. Akan is also with the Center for neXt-generation Communications
(CXC), Department of Electrical and Electronics Engineering, Koç University,
34450 Istanbul, Turkey (email:oba21@cam.ac.uk)}% <-this % stops a space
  }

% ====================================================================
\maketitle

% === ABSTRACT ====================================================================
% =================================================================================
\begin{abstract}
Abstract—Low-Earth-orbit (LEO) inter-satellite links must cope with
strongly doubly selective channels and aged channel state information (CSI).
In this paper, the term ``sensing'' refers to the receiver-side identifiability of a
small set of dominant delay--Doppler path parameters, quantified via CRLB-type proxies,
rather than a full-fledged target-sensing pipeline. Affine frequency division multiplexing (AFDM) provides a sparse delay--Doppler (DD) representation well suited to such channels, yet most existing AFDM designs assume ideal CSI, operate on grid-based channel coefficients, and optimize only communication performance. This paper proposes a two-stage AFDM-based ISAC framework for mobile LEO ISLs that explicitly operates under predicted CSI. In Stage~I, we model the channel by a small number of dominant specular paths and perform sequence prediction directly on their complex gains, delays, and Dopplers, from which we reconstruct the AFDM DD-domain kernel used as the sole instantaneous CSI at the transmitter. In Stage~II, we design a sensing-aware AFDM pre-equalizer by augmenting the classical minimum mean-square error (MMSE) solution with a term obtained from Cram\'er--Rao-type sensitivity measures evaluated under the predicted channel model, leading to a first-order surrogate of a CRLB-regularized pre-equalizer with a single tuning parameter that controls the communication--sensing tradeoff. Simulation results for representative LEO ISL trajectories show that the proposed path-level predictor improves effective-kernel reconstruction over AFDM-unaware baselines, and that, under predicted CSI, the sensing-aware pre-equalizer significantly improves sensing-oriented metrics over outdated-CSI baselines while keeping symbol error rates close to a communication-oriented MMSE design with only modest additional complexity.
\end{abstract}

% === KEYWORDS ====================================================================
% =================================================================================

\IEEEpeerreviewmaketitle

% === I. INTRODUCTION =============================================================
% =================================================================================
\section{Introduction}

%Low-Earth-orbit (LEO) inter-satellite links (ISLs) are emerging as a key component of non-terrestrial networks in 6G, but the associated channels are strongly time- and frequency-selective due to high orbital velocities, large-scale geometry, and sparse specular multipath \cite{xie2021leo,9579443,prol2022position,baeza2022overview}. In such scenarios, classical OFDM waveforms suffer from short coherence times, severe inter-carrier interference, and large cyclic-prefix overhead when the delay spread constitutes a non-negligible fraction of the frame duration. Delay--Doppler (DD) domain waveforms such as orthogonal time--frequency space (OTFS) and affine frequency division multiplexing (AFDM) have therefore attracted considerable attention: by multiplexing data in the DD domain and explicitly representing doubly dispersive channels, they can maintain a nearly time-invariant, sparse channel kernel and enable more robust equalization and sensing in high-mobility regimes \cite{OTFS_survey,OTFS_equalization,AFDM_full_diversity,AFDM_equalization,OTFS_AFDM_comparison,AFDM_survey}. In this work we focus on a single-antenna LEO ISL between two satellites employing AFDM, and we design transmitter-side DD-domain processing that jointly serves communication and sensing objectives.

Low-Earth-orbit (LEO) inter-satellite links (ISLs) are a key component of NTNs in 6G and the backbone of space-based AI infrastructures, where clusters of satellites host ML accelerators interconnected by high-throughput free-space optical ISLs \cite{aguera2025spaceai}. This paper instead focuses on RF/THz LEO ISLs, which face similar requirements for robust, low-latency connectivity. While current operational ISLs are often modeled as purely line-of-sight (LOS), dense satellite formations and in-band monitoring of nearby objects in mega-constellations may introduce a small number of additional sparse specular components, so that the associated channels may become strongly time- and frequency-selective due to high orbital velocities and large-scale geometry \cite{xie2021leo,9579443,prol2022position,baeza2022overview}. In such scenarios, classical OFDM waveforms suffer from short coherence times, severe inter-carrier interference, and large cyclic-prefix overhead when the delay spread constitutes a non-negligible fraction of the frame duration. Delay--Doppler (DD) domain waveforms such as orthogonal time--frequency space (OTFS) and affine frequency division multiplexing (AFDM) have therefore attracted considerable attention: by multiplexing data in the DD domain and explicitly representing doubly dispersive channels, they can maintain a nearly time-invariant, sparse channel kernel and enable more robust equalization and sensing in high-mobility regimes \cite{OTFS_survey,OTFS_equalization,AFDM_full_diversity,AFDM_equalization,OTFS_AFDM_comparison,AFDM_survey}. In this work we focus on a single-antenna LEO ISL between two satellites employing AFDM, and we design transmitter-side DD-domain processing that jointly serves communication and sensing objectives.

A central challenge in this setting is the joint impact of channel state information (CSI) aging and sensing observability. In practical LEO systems, CSI at the transmitter is usually obtained from pilot-based estimation and feedback, together with slowly updated orbital or geometric information, so it can quickly become outdated relative to the AFDM frame duration in strongly time-varying channels. Meanwhile, conventional minimum mean-square error (MMSE) equalization and precoding for multicarrier DD-domain waveforms are optimized solely for symbol detection; they tend to flatten the DD-domain kernel and may inadvertently suppress a small number of dominant paths that are critical for sensing tasks such as delay--Doppler parameter estimation. This motivates AFDM-based ISAC processing that, under predicted path-level CSI at the transmitter, preserves a small number of geometry-consistent dominant paths for sensing while maintaining competitive communication performance.

Several lines of work address parts of this problem, but not the complete picture considered here. First, there is a rich literature on OTFS/AFDM channel modeling, detection, and equalization in doubly dispersive channels, including low-complexity MMSE detectors and message-passing receivers that exploit the sparse DD-domain representation \cite{OTFS_equalization,AFDM_full_diversity,AFDM_equalization,OTFS_AFDM_comparison}. These works typically assume perfect or near-perfect instantaneous CSI and focus on bit- or symbol-error performance. Second, deep learning-based CSI predictors using recurrent networks, Transformers, or graph neural networks have been proposed for high-mobility channels, but they generally operate on grid-based CSI (e.g., time--frequency or space--frequency coefficients) and are evaluated mainly using NMSE, without tight coupling to specific DD-domain waveforms or sensing metrics \cite{luo2018channel,zhang2021deep,zhang2022,ying2024deep,10637286,DL_CSI_pred_survey}. Third, ISAC waveform and precoder designs often formulate joint communication--sensing objectives using Fisher information and Cram\'er--Rao lower bounds (CRLBs), but rely on simplified analytic channel models or ideal CSI and are typically developed in OFDM or narrowband MIMO frameworks rather than AFDM-type DD waveforms \cite{ISAC_CRLB_survey,wei2023integrated,dong2024debrisense}. In the specific context of LEO THz
satellite channels, fundamental limits of cooperative ISAC have been characterized using unified ISAC signal models and normalized CRLB expressions that explicitly account for hardware impairments, phase noise, and pointing errors \cite{fundamental_coop_thz_isac}.

In this paper, we take a step toward closing this gap for a mobile LEO inter-satellite link with AFDM modulation by proposing a two-stage AFDM–ISAC architecture that is explicitly driven by predicted CSI rather than ideal or fully up-to-date estimates. Stage I adopts a sparse multipath model with line-of-sight and a small number of specular components, tracks and extrapolates a fixed set of dominant delay–Doppler paths with consistent identities, and reconstructs the AFDM delay–Doppler kernel from their predicted complex gains, delays, and Dopplers. Stage II takes this predicted kernel as the sole instantaneous CSI and designs a sensing-aware AFDM pre-equalizer that starts from the classical MMSE solution and introduces a Fisher-information-based sensitivity map as a regularization term; the resulting filter has a first-order closed-form that serves as a computationally efficient surrogate for CRLB-regularized design, with a single scalar weight controlling the tradeoff between symbol MSE and delay–Doppler sensing CRLB. Existing OTFS/AFDM-based ISAC and CRLB-driven designs typically assume perfect or near-perfect instantaneous CSI and operate on grid-based channel representations, without addressing transmitter-side DD-domain pre-equalization under predicted CSI or its coupling with path-level prediction and AFDM delay--Doppler kernel reconstruction.

Together, the two stages form a tightly coupled pipeline: the predictor is trained on the AFDM delay–Doppler kernel used by the pre-equalizer, which always runs under reconstructed predicted CSI rather than idealized channels. This lets us experimentally quantify, for doubly selective LEO ISLs, how prediction errors and kernel reconstruction affect the achievable communication–sensing tradeoff, where ``sensing'' is interpreted as CRLB-based identifiability of dominant delay--Doppler path parameters.

In summary, this paper makes the following contributions.
\begin{itemize}
  \item A path-level channel prediction framework tailored to AFDM-based LEO inter-satellite links. Instead of predicting grid-based CSI taps on time--frequency grids, the predictor tracks a small set of geometry-consistent delay--Doppler paths and reconstructs the AFDM kernel on the delay--Doppler grid, with losses and evaluation metrics that emphasize phase accuracy and path identity in addition to NMSE.
  \item A CRLB-aware AFDM pre-equalizer that explicitly accounts for predicted CSI. A normalized Fisher-information-based sensitivity map perturbs the classical MMSE solution on the AFDM delay--Doppler grid, yielding a low-complexity pre-equalizer with a single scalar that controls the tradeoff between symbol MSE and delay--Doppler sensing CRLB.
  \item A numerical study of the proposed two-stage AFDM--ISAC architecture on doubly selective LEO ISLs. The evaluation compares predicted versus outdated CSI, and different sensitivity maps and normalizations, and identifies operating regimes where delay--Doppler CRLBs improve markedly while symbol error rates remain close to a communication-oriented MMSE baseline.
\end{itemize}

%Our prior work PIST introduced a physics-informed CSI prediction framework for debris-aware LEO-NTN/ISL links \cite{PIST}. PIST constructs a geometry- and debris-aware narrowband channel model, uses ISAC range and radial-velocity measurements of a few dominant occluders as additional inputs and as a sensing-consistency regularizer, and focuses on one-step prediction of complex narrowband CSI time series. It does not address delay–Doppler-domain waveforms or pre-equalizer design. In contrast, the present paper targets AFDM-based integrated sensing and communication: we explicitly learn delay–Doppler paths, reconstruct AFDM kernels, and design a CRLB-aware AFDM pre-equalizer that leverages predicted AFDM kernels to trade communication performance against delay–Doppler sensing accuracy under predicted CSI. At a higher level, such mechanisms could be combined with emerging Internet-of-Space and semantic communication architectures to support task-oriented sensing and delivery \cite{semantic_ios}.
Our prior work PIST \cite{PIST} studied physics-informed narrowband CSI prediction for debris-aware LEO links; here we instead predict delay–Doppler paths, reconstruct AFDM kernels, and design a CRLB-aware AFDM pre-equalizer that trades communication performance against delay–Doppler sensing accuracy under predicted CSI.

The remainder of this paper is organized as follows. Section~II reviews related work on DD-domain waveforms, channel prediction, and ISAC precoding. Section~III presents the LEO ISL scenario, AFDM frame structure, and channel model, and defines the CSI modes considered in this work. Section~IV describes the Stage~I path-level prediction framework and AFDM kernel reconstruction. Section~V develops the Stage~II sensing-aware AFDM pre-equalizer driven by CRLB considerations. Section~VI provides numerical results, tradeoff analysis, and ablation studies. Section~VII concludes the paper and outlines future research directions.

\section{Related Work}

\subsection{DD-domain waveforms and high-mobility ISAC}

DD-domain waveforms such as OTFS have been widely investigated as alternatives to OFDM in high-mobility scenarios because they map time-varying channels into nearly time-invariant DD-domain kernels with a small number of dominant taps, which facilitates robust equalization and, for sensing applications, a natural representation of target range--Doppler pairs \cite{OTFS_survey,OTFS_equalization}. Many works have developed channel estimation and detection methods for OTFS, including message-passing and iterative equalization techniques that exploit the sparse DD-domain representation \cite{OTFS_equalization}. More recently, AFDM has been proposed as a chirp-based multicarrier waveform tailored to doubly dispersive channels, relying on the discrete affine Fourier transform to avoid overlap among delays and Dopplers in an AFDM-domain impulse response and to achieve full diversity in linear time-varying channels \cite{AFDM_full_diversity,AFDM_equalization,AFDM_survey,OTFS_AFDM_comparison}.

On the ISAC side, several works have combined OTFS with radar sensing, leveraging the DD-domain structure to jointly support communications and delay--Doppler radar estimation, and investigating communication--sensing tradeoffs in terms of rate and estimation accuracy \cite{MIMO_OTFS_ISAC}. AFDM-based ISAC designs have also been explored recently, for example in monostatic and bistatic sensing configurations and in system-level AFDM–ISAC architectures \cite{AFDM_ISAC_letter,AFDM_ISAC_bistatic,AFDM_ISAC_framework}. These studies typically derive signal models and detection algorithms for OTFS- or AFDM-based radar and focus on receiver-side processing. They generally assume perfect or high-quality channel knowledge at the transmitter or receiver. There appears to be limited work on transmitter-side DD-domain pre-equalization for AFDM or OTFS that explicitly accounts for sensing metrics and operates under predicted CSI in high-mobility LEO ISLs, which is the regime targeted in this paper.

\subsection{Channel prediction and physics-informed learning}

Deep learning-based channel prediction has been extensively studied for terrestrial and non-terrestrial systems. Recurrent neural networks, Transformers, and graph neural networks have been employed to predict future CSI samples on time--frequency or space--frequency grids, mainly targeting NMSE improvements and robustness to pilot overhead reduction \cite{luo2018channel,zhang2021deep,zhang2022,ying2024deep,10637286,DL_CSI_pred_survey}. Some works incorporate geometric or physical priors, such as angle--delay sparsity or parametric channel models, into the learning procedure, for example by constraining the representation or adding regularization terms inspired by physical consistency \cite{10599118,wagle2025physics}. Nevertheless, most of these methods still operate on grid-based CSI and are not directly structured to output path parameters that can be naturally consumed by DD-domain waveform designs, and therefore are not directly suited to support the sparse DD-domain ISAC pre-equalization proposed here.

%Our prior work PIST \cite{PIST} is a representative physics-informed CSI predictor for debris-aware LEO-NTN/ISL links. Starting from TLE-driven orbital geometry and a debris-aware narrowband channel model with a line-of-sight component and a small number of specular debris-induced paths, PIST generates geometry-consistent complex CSI time series. A lightweight convolutional–Transformer encoder performs one-step prediction of this narrowband CSI, while integrated sensing provides noisy range and radial-velocity measurements of a few dominant occluders. These observables are used both as additional inputs and through a differentiable sensing-consistency loss between ISAC measurements and CSI-implied kinematics, improving prediction quality and robustness compared with purely data-driven baselines. However, PIST and related approaches stop at the level of improved CSI sequences; they do not output delay–Doppler path parameters or AFDM kernels, nor do they design waveform-level pre-equalizers that explicitly balance communication and sensing performance.

%In contrast, the Stage~I design in this paper performs prediction directly at the path-parameter level and reconstructs an AFDM delay--Doppler kernel from a small set of dominant paths, providing a path-based interface that can be directly consumed by the CRLB-aware AFDM pre-equalizer in Stage~II.

Our prior work PIST \cite{PIST} is a representative physics-informed CSI predictor for debris-aware LEO-NTN/ISL links. It uses a geometry- and debris-aware narrowband channel model, a lightweight convolutional--Transformer encoder, and sensing-consistency regularization to improve one-step prediction of complex narrowband CSI, but still operates on grid-based narrowband CSI sequences. PIST and related approaches do not output delay--Doppler path parameters or AFDM kernels, nor do they design waveform-level pre-equalizers that explicitly balance communication and sensing performance. In contrast, the Stage~I design in this paper performs prediction directly at the path-parameter level and reconstructs an AFDM delay--Doppler kernel from a small set of dominant paths as a path-based interface for the CRLB-aware AFDM pre-equalizer in Stage~II.

\subsection{ISAC precoding and tradeoff design}

A large body of ISAC research has focused on precoding and waveform design that jointly optimize communication and sensing metrics, often using Fisher information and CRLBs as sensing performance measures. In multi-antenna OFDM and wideband systems, for example, precoders have been designed to minimize CRLB-based localization or direction-of-arrival errors under communication rate constraints, or to balance communication SINR against beampattern mismatch \cite{ISAC_CRLB_survey,wei2023integrated}. These works typically assume relatively simple analytic channel models and perfect or slowly varying CSI, and they often restrict attention to narrowband or frequency-flat approximations when deriving CRLB expressions and optimization problems.

More recent studies on cooperative ISAC networks and multi-cell ISAC explore CRLB scaling laws, coordinated beamforming, and power allocation across distributed transceivers, again under idealized CSI assumptions \cite{Coop_ISAC_survey}. While these results provide valuable insights into network-level ISAC tradeoffs, they do not directly address waveform-level pre-equalization in strongly doubly dispersive channels, nor do they consider the impact of prediction-driven CSI on precoder design. To the best of our knowledge, there is no prior work that combines path-level prediction of a doubly selective LEO ISL, AFDM delay--Doppler kernel reconstruction, and a sensing-aware pre-equalizer derived from CRLB considerations and evaluated under predicted CSI. Prior ISAC precoding frameworks typically evaluate CRLBs under perfect CSI and do not explicitly account for CSI aging or prediction errors, nor do they target AFDM-based ISLs where DD-domain kernels must be reconstructed from predicted path parameters. This three-way combination is precisely the regime targeted by the proposed two-stage AFDM--ISAC framework.

\section{System Model and Performance Metrics}

\subsection{LEO inter-satellite link and delay–Doppler frame structure}

% We consider a single-antenna LEO inter-satellite link (ISL)
% between a transmitter satellite and a receiver satellite. Due to
% the high orbital velocity and the coexistence of a line-of-sight
% path with several scattered paths, the channel within one frame
% is doubly selective, with finite delay spread and Doppler spread.
% To obtain a stable and sparse representation under such
% conditions, we adopt an AFDM frame structure in the
% delay–Doppler (DD) domain. The considered LEO ISL geometry and AFDM frame are illustrated in Fig.~\ref{fig:system} and should be interpreted as a simplified link-level abstraction for a single representative ISL; constellation-level scheduling, detailed orbital perturbations, hardware impairments, and non-specular clutter are not explicitly modeled and are instead absorbed into the path parameters and effective noise.

\begin{figure}[!t]
  \centering
  \includegraphics[width=0.85\columnwidth]{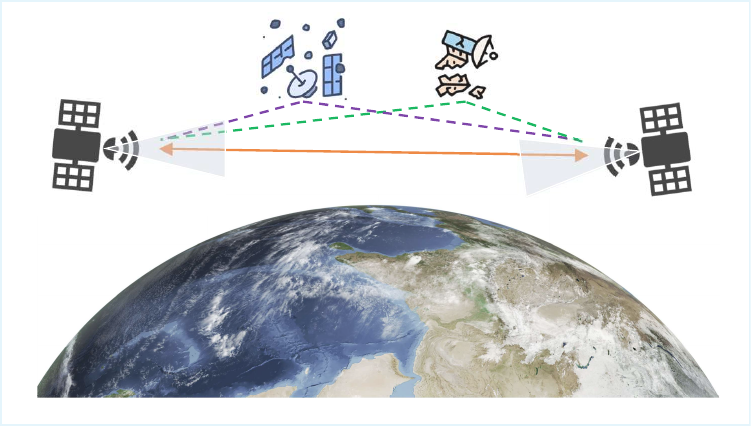}
  \caption{Geometry of the considered LEO inter-satellite link with debris-induced LOS/NLOS paths.}
  \label{fig:system}
\end{figure}

We consider a single-antenna AFDM-based LEO ISL between a transmitter and a receiver and adopt a delay–Doppler AFDM frame to obtain a sparse channel representation. The geometry and frame in Fig.~\ref{fig:system} serve as a simplified link-level abstraction for one representative ISL; constellation-level scheduling, detailed orbital perturbations, hardware impairments, and non-specular clutter are absorbed into the path parameters and effective noise.

% Fig.1: 几何场景 / 系统模型

Each AFDM frame occupies M delay bins and N Doppler
bins on a DD grid with coordinates
\[
(\ell \Delta\tau, k \Delta\nu), \quad \ell = 0,\ldots,M-1,\; k = 0,\ldots,N-1,
\]
where $\Delta\tau$ and $\Delta\nu$ are the delay and Doppler
sampling intervals determined by the system bandwidth,
symbol period, and the maximum delay and Doppler spreads
of the LEO ISL. The transmitter maps a complex symbol
matrix $X[\ell,k]$ on this DD grid to a time-domain waveform
via AFDM modulation, while the receiver applies a matched
filter and inverse AFDM transform to obtain the DD-domain
observation $Y[\ell,k]$. To enforce a consistent power convention
across frames, we assume
\[
\frac{1}{MN} \sum_{\ell=0}^{M-1} \sum_{k=0}^{N-1}
\mathbb{E}\!\left[ |X[\ell,k]|^2 \right] = E_s,
\]
where $E_s$ denotes the nominal symbol energy. All delay–Doppler
channel kernels, precoders, and performance metrics in the
remainder of the paper are defined with respect to this AFDM
DD grid and power normalization.

\subsection{Sparse multipath channel and DD representation}

Within one AFDM frame, the LEO ISL is modeled as a finite sum of
specular paths on top of thermal noise. The continuous-time baseband
time-varying impulse response is written as
\begin{equation}
  h(t,\tau)=\sum_{p=1}^{P}\alpha_p e^{j\phi_p} e^{j2\pi \nu_p t}\,\delta(\tau-\tau_p),
\end{equation}
where, for path \(p\), \(\alpha_p\ge 0\) and \(\phi_p\in[-\pi,\pi)\) denote
the amplitude and phase, \(\tau_p\) is the propagation delay, and \(\nu_p\) is
the Doppler shift induced by the relative motion between the satellite and
the partner satellite. The received baseband signal is
\begin{equation}
  y(t)=\int h(t,\tau)\,x(t-\tau)\,d\tau+w(t).
\end{equation}
In the considered LEO ISL geometry, only a few dominant clusters contribute appreciably within one AFDM frame; hence we adopt a sparse path model with parameters $(\tau_p,\nu_p)$ for link-level evaluation.

After AFDM modulation, propagation through the continuous-time channel with impulse response $h(t,\tau)$, and the receive-side 
matched filtering and discretization described above, the channel is 
represented on the DD grid by a complex-valued kernel
\begin{equation}
  H[\ell,k;t] \in \mathbb{C}^{M\times N}, \qquad 0 \le \ell < M,\; 0 \le k < N,
\end{equation}
whose samples describe how transmitted symbols placed around \((\ell,k)\) in 
the DD plane are spread across neighboring grid points. The mapping from the 
continuous path parameters \(\{\alpha_p,\phi_p,\tau_p,\nu_p\}\) to the 
discrete kernel \(H[\ell,k;t]\) follows the AFDM synthesis used in our 
implementation; the same mapping is applied to both measured and predicted 
path parameters and need not be repeated here.

For notational convenience, we collect all path-level quantities at time \(t\) 
into
\begin{equation}
  \Theta(t) = \big\{ \alpha_p(t), \phi_p(t), \tau_p(t), \nu_p(t) \big\}_{p=1}^{P},
\end{equation}
and view \(H[\ell,k;t]\) as a deterministic function of \(\Theta(t)\) and of 
the fixed AFDM grid \((M,N,\Delta_\tau,\Delta_\nu)\).

\subsection{DD-domain input--output model and SNR}

On the AFDM DD grid, the per-frame input--output relation is modeled as
\begin{equation}
  Y[\ell,k] = (H \circledast X)[\ell,k] + W[\ell,k],
\end{equation}
where $\circledast$ denotes two-dimensional circular convolution on the $M \times N$
DD grid (indices taken modulo $(M,N)$), and $W[\ell,k] \sim \mathcal{CN}(0,\sigma_w^2)$
is i.i.d. complex Gaussian noise.
The corresponding two-dimensional DFT representation will be used later for 
precoder design and is not repeated here.

With the power convention above, the average symbol power per frame is 
\(E_s\). We define the nominal signal-to-noise ratio as
\begin{equation}
  \mathrm{SNR} = \frac{E_s}{\sigma_w^2},
\end{equation}
and use this as the reference SNR in all simulations.

\subsection{Effective channel, CSI modes, and performance metrics}

In the DD domain, the precoder can be viewed as a two-dimensional filter 
applied to the symbol matrix \(X[\ell,k]\). Given some available CSI for frame 
\(t\) and a scalar tradeoff weight \(\lambda_{\mathrm{ISAC}}\), the transmitter 
constructs a DD-domain precoding kernel \(G^{(\lambda)}[\ell,k]\) and embeds 
it into the AFDM modulation. For a fixed physical channel kernel \(H[\ell,k]\) 
within a frame and the corresponding precoder kernel \(G^{(\lambda)}[\ell,k]\), 
the effective DD-domain channel is defined as
\begin{equation}
  H_{\mathrm{eff}}^{(\lambda)}[\ell,k]
  = (H \circledast G^{(\lambda)})[\ell,k].
\end{equation}
This kernel captures the overall channel seen after AFDM demodulation, 
including both the propagation channel and the precoder.

We consider two CSI modes under this model. In the predicted-CSI mode, the 
precoder is designed from a channel kernel \(\hat{H}_t[\ell,k]\) obtained by 
mapping path-parameter predictions for frame \(t\) onto the AFDM DD grid. In 
the outdated-CSI baseline, the precoder is instead designed from 
\(H_{t-1}[\ell,k]\), the kernel available from the previous frame, while the 
actual propagation channel is still \(H_t[\ell,k]\). In both cases, the AFDM 
configuration, transmit power, noise variance, and DD-domain input--output 
model are identical; only the timeliness of the CSI used to construct 
\(G^{(\lambda)}[\ell,k]\) differs.

We evaluate communication by per-frame symbol MSE and SER, and sensing by CRLB-based metrics on key path parameters (e.g., delays and Dopplers); detailed definitions are provided in the subsequent sections.

\section{Stage I: Path-Level Prediction and AFDM Kernel Reconstruction}
\label{sec:stage1}

\subsection{Minimal channel abstraction}

\begin{figure*}[!t]
  \centering
  \includegraphics[width=0.9\textwidth]{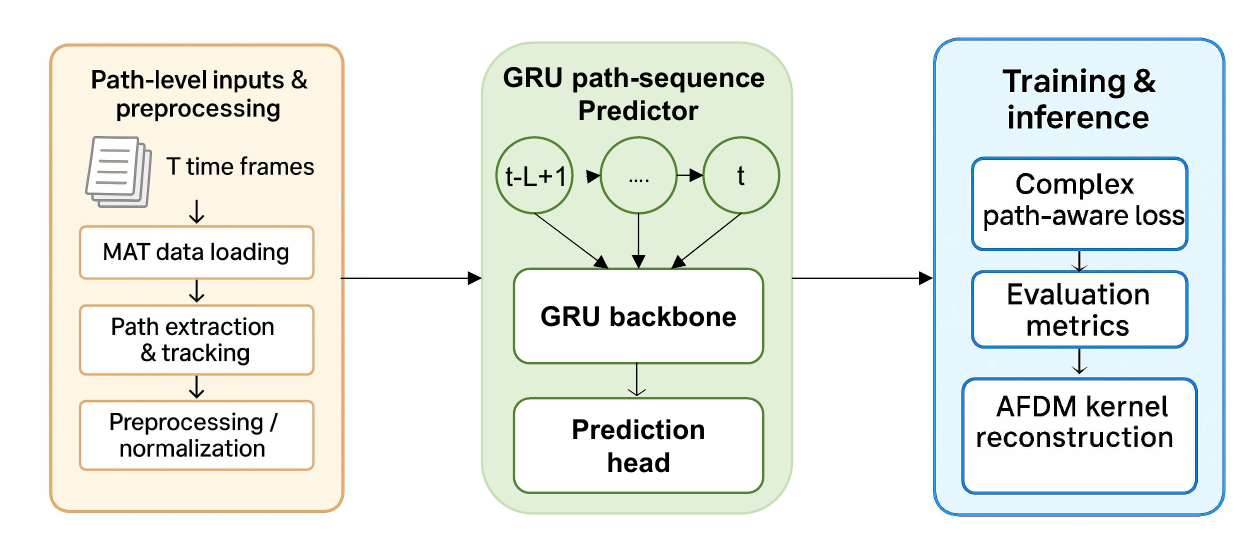}
  \caption{Stage I path-level CSI prediction and AFDM kernel reconstruction pipeline.}
  \label{fig:stage1}
\end{figure*}

Under the AFDM DD frame and power convention in Section~III, 
we adopt a geometry-driven multi-path model with line-of-sight 
and single-bounce components. At each discrete time t, each path 
p is parameterized by its complex gain $h_p(t)$, delay $\tau_p(t)$, 
and Doppler $\nu_p(t)$. For Stage~I, all AFDM- and grid-related 
quantities ($\Delta_\tau, \Delta_\nu, M, N$) are treated as fixed 
system parameters determined at design time; the predictor operates 
purely on the path-parameter sequences $\{h_p(t), \tau_p(t), \nu_p(t)\}$ 
and does not modify the underlying DD lattice. The overall Stage-I pipeline is summarized in Fig.~\ref{fig:stage1}

% Fig.2: Stage I 路径级预测框架

\subsection{Dominant-path selection and identity tracking}
We keep $K$ dominant paths per $t$ by sorting $|h_p(t)|$ and retaining the top-$K$ tuples $\{h_k,\tau_k,\nu_k\}_{k=1}^K$.
To prevent index swapping, we match dominant paths between $t-1$ and $t$ using a normalized delay–Doppler distance and solve a one-to-one assignment.
The matching cost is a weighted sum of normalized delay and Doppler deviations, and candidates beyond a fixed distance threshold are treated as unmatched.
Unmatched slots are filled with zero-amplitude placeholders so that appearances and disappearances are explicit.
This yields temporally consistent sequences $\{A_k,\theta_k,\tau_k,\nu_k\}$ with $A_k=|h_k|$ and $\theta_k=\arg h_k$.

\subsection{Feature encoding and predictor I/O}
To stabilize learning, amplitudes are floored at $A_{\min}$ and mapped as $a_k(t)=\log(\max(A_k(t),A_{\min}))$; phase uses $s_k(t)=\sin\theta_k(t)$ and $c_k(t)=\cos\theta_k(t)$.
The set $\{a_k,\tau_k,\nu_k\}$ is standardized using training-split statistics $(\mu,\sigma)$ and then frozen.
At each $t$, we form $x(t)\in\mathbb{R}^{5K}$ by concatenating
$[\tilde a_{1:K},\, s_{1:K},\, c_{1:K},\, \tilde\tau_{1:K},\, \tilde\nu_{1:K}]$.
A window $X(t_0)\in\mathbb{R}^{L\times 5K}$ is fed into a multi-layer GRU of hidden size $d$; the last state is projected to $z\in\mathbb{R}^{d'}$ and mapped by a linear head to $\hat Y(t_0)\in\mathbb{R}^{H\times 5K}$ for horizon $H$ (we use $H=1$ in experiments).
Causality is enforced by construction. Throughout the experiments we set the context length to \(L=64\), matching the one-step prediction format used across all architectural baselines.

\subsection{Training objective and loss design}
The loss is a weighted sum
\[
L
=
w_A L_A
+
w_\theta L_\theta
+
w_\tau L_\tau
+
w_\nu L_\nu
+
w_{\mathrm{uc}} L_{\mathrm{uc}}
+
w_{\mathrm{c}} L_{\mathrm{cnmse}}.
\]
$L_A$ is MSE on log-amplitude; $L_\theta$ is amplitude-weighted MSE on $(s_k,c_k)$; $L_\tau$ and $L_\nu$ are Huber losses on standardized delay/Doppler; $L_{\mathrm{uc}}$ penalizes $(\hat s_k^2+\hat c_k^2-1)^2$; $L_{\mathrm{cnmse}}$ enforces complex-plane consistency via normalized $|\hat h_k-h_k|^2$ with $\hat h_k=\exp(\hat a_k)(\hat c_k+\mathrm{j}\hat s_k)$.
All statistics are computed on the training split; chronological splits include guard intervals to avoid leakage.
We fix $(w_A,w_\theta,w_\tau,w_\nu,w_{\mathrm{uc}},w_{\mathrm{c}})$ to a single configuration selected by a one-time sweep on the training/validation split and reuse this setting across all geometries.
Training uses mini-batches of size $B$, AdamW with initial learning rate $10^{-3}$ and cosine decay, gradient clipping on the global norm, and early stopping on validation NMSE.

\subsection{AFDM kernel reconstruction and Stage-II interface}
For a predicted time $t$, we de-normalize to obtain $\hat A_k$, $\hat\theta_k$, $\hat\tau_k$, and $\hat\nu_k$, and form $\hat h_k=\hat A_k e^{\mathrm{j}\hat\theta_k}$.
Using the DD grid $\{\tau_\ell\}_{\ell=0}^{M-1}$ and $\{\nu_n\}_{n=0}^{N-1}$ 
with resolutions ($\Delta_\tau,\Delta_\nu$) defined in Section~III, we define fractional indices $\hat\ell_k=\hat\tau_k/\Delta\tau$ and $\hat n_k=\hat\nu_k/\Delta\nu$.
The predicted delay–Doppler kernel is
\[
\hat H[\ell,n;t]
=
\sum_{k=1}^K \hat h_k\,
D_M\!(\tfrac{\ell-\hat\ell_k}{M})\,
D_N\!(\tfrac{n-\hat n_k}{N})\,
\exp(-\mathrm{j}\,2\pi n\,\tfrac{\hat \ell_k}{N}),
\]
where $D_M$ and $D_N$ are Dirichlet kernels and the exponential is the AFDM phase-coupling term.
This yields the Stage-I output $H_{\mathrm{pred}}(t)$, which is the sole CSI input to Stage-II pre-equalization. For evaluation, we apply the same AFDM mapping to ground-truth path parameters at $t$ and $(t-\Delta t)$ to form the reference and outdated-CSI baselines (Section~VI).

\subsection{Complexity and implementation note}
Per step, path tracking and encoding are $O(K)$; GRU inference is $O(Ld^2)$; kernel synthesis is $O(KMN)$ but is implemented with precomputed Dirichlet tables and FFT-accelerated convolution to near $O((K+1)MN)$.
All Stage-I hyperparameters $(K,L,d,d')$ and loss weights are fixed once and reused for all reported results, so that performance differences arise only from the CSI source (predicted vs outdated) and from the Stage-II design.

\section{Stage II: CRLB-Aware AFDM Pre-Equalization under Predicted CSI}
\label{sec:stage2}

\subsection{AFDM-domain model and MMSE baseline}

Stage~II operates on the AFDM delay–Doppler lattice defined in Section~III, using the AFDM-domain kernels $H[\ell,n;t]\in\mathbb C^{M\times N}$ constructed in Stage I from the chosen CSI source (predicted or outdated).

Given one AFDM frame $X[\ell,n]\in\mathbb C^{M\times N}$, the input–output relation without pre-equalization is modeled as a two-dimensional circular convolution
\begin{equation}
Y[\ell,n]
=
(H\circledast X)[\ell,n;t]
+
W[\ell,n],
\end{equation}
where indices are taken modulo $(M,N)$ and $W[\ell,n]\sim\mathcal{CN}(0,\sigma_w^2)$.
Let $\mathcal F_2\{\cdot\}$ denote the 2D discrete Fourier transform (2D-DFT).
In the frequency domain,
\begin{align}
X_f[m,n] &= \mathcal F_2\{X[\ell,n]\},\\
H_f[m,n;t] &= \mathcal F_2\{H[\ell,n;t]\},\\
Y_f[m,n] &= H_f[m,n;t]\,X_f[m,n] + W_f[m,n],
\end{align}
so the channel decouples across $(m,n)$ and scalar filters can be designed per bin.

A transmit pre-equalizer $G[\ell,n;t]$ induces the effective kernel
\begin{equation}
H_{\mathrm{eff}}[\ell,n;t]
=
(H\circledast G)[\ell,n;t],
\end{equation}
or, in the frequency domain,
\begin{equation}
H_{\mathrm{eff},f}[m,n;t]
=
H_f[m,n;t]\,G_f[m,n;t],
\end{equation}
with $G_f[m,n;t]=\mathcal F_2\{G[\ell,n;t]\}$.

Ignoring sensing, a classical MMSE design treats each bin as a complex AWGN channel with noise variance $\sigma_w^2$ and i.i.d.\ data of average power $E_s$, and defines $\gamma=\sigma_w^2/E_s$.
The per-bin MMSE pre-equalizer then reads
\begin{equation}
G_f^{\mathrm{mmse}}[m,n;t]
=
\alpha(t)\,
\frac{H_f^*[m,n;t]}
     {|H_f[m,n;t]|^2 + \gamma},
\label{eq:preq13}
\end{equation}
where $\gamma = \sigma_w^2 / E_s$ and $\alpha(t)$ is a real-valued scaling factor. In practice we first compute the unscaled Wiener filter with $\alpha(t)=1$, take its inverse 2D-DFT to obtain $G^{\mathrm{mmse}}[\ell,n;t]$, and then rescale it so that the average transmit power per frame
\[
P_G(t)
=
\frac{1}{MN}
\sum_{\ell,n}
|G^{\mathrm{mmse}}[\ell,n;t]|^2
\]
matches a target symbol power $E_s$. The effective kernel
\[
H_{\mathrm{eff}}^{\mathrm{mmse}}[\ell,n;t]
=
(H \circledast G^{\mathrm{mmse}})[\ell,n;t]
\]
is further normalized to have unit average squared magnitude over $(\ell,n)$, which fixes the effective SNR when comparing different CSI sources and different values of the sensing weight in the numerical results. The same two-step normalization of $G$ and $H_{\mathrm{eff}}$ is applied when $G_f^{\mathrm{mmse}}$ is replaced by the ISAC-regularized pre-equalizer.

\subsection{Sensing metric and joint objective}

Besides symbol detection, Stage~II encourages good identifiability of a small set of dominant paths, using Fisher/CRLB expressions as a design proxy.
Let $\{h_k,\tau_k,\nu_k\}_{k=1}^{K'}$ denote the $K'\leq K$
strongest delay--Doppler paths at snapshot $t$, and collect
their real-valued parameters into
\begin{equation}
\boldsymbol{\eta}
=
\{
\Re\{h_k\},\Im\{h_k\},\tau_k,\nu_k
\}_{k=1}^{K'}
\in \mathbb{R}^{P},
\label{eq:eta_vec}
\end{equation}
with $P = 4K'$ in our implementation.
For a fixed probing frame $X[\ell,n]$ and pre-equalizer
$G[\ell,n;t]$, the stacked frequency-domain observation
$\mathbf{y}\in\mathbb{C}^{MN}$ can be written as
\begin{equation}
\mathbf{y}
=
\boldsymbol{\mu}(\boldsymbol{\eta};G)
+
\mathbf{w},
\quad
\mathbf{w}\sim\mathcal{CN}(\mathbf{0},\sigma_w^2\mathbf{I}),
\label{eq:y_mu}
\end{equation}
where $\boldsymbol{\mu}(\boldsymbol{\eta};G)$ is the noise-free
mean induced by the AFDM effective kernel and pre-equalizer.

For this complex Gaussian model, the Fisher information
matrix (FIM) of $\boldsymbol{\eta}$ is given by the Slepian--Bangs
formula
\begin{equation}
\mathbf{I}(\boldsymbol{\eta};G)
=
\frac{2}{\sigma_w^2}
\Re\{
\mathbf{J}(\boldsymbol{\eta};G)^{H}
\mathbf{J}(\boldsymbol{\eta};G)
\},
\label{eq:fim}
\end{equation}
where
$\mathbf{J}(\boldsymbol{\eta};G)
=
\partial\boldsymbol{\mu}(\boldsymbol{\eta};G)/
\partial\boldsymbol{\eta}^{\mathsf{T}}$
is the Jacobian.
The associated CRLB matrix is
$\mathbf{C}(\boldsymbol{\eta};G) = \mathbf{I}(\boldsymbol{\eta};G)^{-1}$,
and any unbiased estimator $\widehat{\boldsymbol{\eta}}$ satisfies
$\mathrm{cov}(\widehat{\boldsymbol{\eta}})\succeq
\mathbf{C}(\boldsymbol{\eta};G)$.

As a scalar sensing metric we use the trace of the CRLB,
\begin{equation}
J_{\mathrm{sense}}(G)
=
\mathrm{tr}
(
\mathbf{C}(\boldsymbol{\eta};G)
),
\label{eq:Jsense_def}
\end{equation}
which measures the aggregate variance of the dominant-path
parameters.
In the AFDM setting,
$\boldsymbol{\mu}(\boldsymbol{\eta};G)$ involves Dirichlet-type
pulse shaping, circular convolution, and a 2D DFT, so that a
closed-form $\mathbf{J}(\boldsymbol{\eta};G)$ would be cumbersome;
in practice we evaluate $\mathbf{J}$ and $\mathbf{I}$ numerically
via small finite-difference perturbations of $\boldsymbol{\eta}$.
Since Stage~II operates under predicted CSI, the FIM and
CRLB are evaluated with respect to the predicted AFDM kernel
and should be viewed as relative sensing figures of merit
under this model.
In the numerical results we report the normalized metric
$J_{\mathrm{sense}}(G)/J_{\mathrm{sense}}(G^{\mathrm{mmse}})$,
denoted as $\mathrm{CRLB}(\lambda)/\mathrm{CRLB}(0)$.

On the communication side we use the standard symbol MSE
under linear MMSE detection as the cost $J_{\mathrm{comm}}(G_f)$,
whose minimizer has the form in \eqref{eq:preq13}.
Introducing a nonnegative weight $\lambda_{\mathrm{ISAC}}\geq 0$,
we combine both objectives as
\begin{equation}
\min_{G_f}
\;
J_{\mathrm{comm}}(G_f)
+
\lambda_{\mathrm{ISAC}}\,
J_{\mathrm{sense}}(G_f),
\label{eq:joint_obj}
\end{equation}
where $G_f$ denotes the 2D DFT of $G[\ell,n;t]$ and
$J_{\mathrm{sense}}(G_f)$ is evaluated via
\eqref{eq:Jsense_def} using the corresponding AFDM kernel.

\subsection{Approximate closed-form ISAC pre-equalizer}

Solving \eqref{eq:joint_obj} exactly would require repeatedly
constructing and inverting the FIM for each candidate
pre-equalizer and is computationally demanding.
To obtain a simple design, we adopt a first-order local
approximation of the sensing cost around the MMSE solution.

For each bin $(m,n)$ and snapshot $t$, let
\begin{equation}
q_{m,n}
=
|
H_f[m,n;t]\,
G_f[m,n;t]
|^2
\label{eq:q_def}
\end{equation}
denote the post-equalization power on that bin, and write
$q^{\mathrm{mmse}}_{m,n}$ for the value induced by
$G^{\mathrm{mmse}}_f[m,n;t]$ in \eqref{eq:preq13}.
By the chain rule,
\begin{equation}
\frac{\partial J_{\mathrm{sense}}}{\partial q_{m,n}}
=
\mathrm{tr}
(
\frac{\partial \mathbf{C}}{\partial q_{m,n}}
),
\label{eq:mmse-preq}
\end{equation}
and increasing $q_{m,n}$ on informative bins tends to decrease
$J_{\mathrm{sense}}(G_f)$ in standard Gaussian models.
We therefore define the local sensitivity
\begin{equation}
S_f[m,n;t]
\triangleq
-
\left.
\frac{\partial J_{\mathrm{sense}}(G_f)}{\partial q_{m,n}}
\right|_{G_f = G^{\mathrm{mmse}}_f},
\label{eq:Sf_def}
\end{equation}
so that larger $S_f[m,n;t]$ indicates that allocating additional
post-equalization power on bin $(m,n)$ is more effective in
reducing the CRLB metric.
A first-order Taylor expansion of $J_{\mathrm{sense}}(G_f)$ around
$G^{\mathrm{mmse}}_f$ yields
\begin{align}
J_{\mathrm{sense}}(G_f)
&\approx
J_{\mathrm{sense}}(G^{\mathrm{mmse}}_f)
+
\sum_{m,n}
\frac{\partial J_{\mathrm{sense}}}{\partial q_{m,n}}
(
q_{m,n} - q^{\mathrm{mmse}}_{m,n}
)
\nonumber\\
&\approx
\text{const}
-
\sum_{m,n}
S_f[m,n;t]\,
q_{m,n},
\label{eq:Jsense_firstorder}
\end{align}
where terms independent of $G_f$ have been absorbed into
the constant.
Substituting \eqref{eq:q_def} gives the surrogate
\begin{align}
\widetilde{J}_{\mathrm{sense}}(G_f)
&= - \sum_{m,n}
S_f[m,n; t]\,
|
H_f[m,n; t]\,G_f[m,n; t]
|^2.
\label{eq:Jsense_surrogate}
\end{align}

Using \eqref{eq:Jsense_surrogate}, the joint objective
\eqref{eq:joint_obj} is approximated (up to constants) by
\begin{align}
J_{\mathrm{tot}}(G_f)
&\approx J_{\mathrm{comm}}(G_f)
\nonumber\\
&\quad
- \lambda_{\mathrm{ISAC}}
\sum_{m,n}
S_f[m,n; t]\,
|
H_f[m,n; t]\,
G_f[m,n; t]
|^2.
\label{eq:Jtot_surrogate}
\end{align}

Following the usual scalar AWGN model per bin, we write
the per-bin communication MSE as
\begin{equation}
J_{\mathrm{comm}}(G_f)
=
\sum_{m,n}
(
|1 - g H_f[m,n;t]|^2
+
\gamma |g|^2
),
\end{equation}
with $\gamma = \sigma_w^2/E_s$ as in \eqref{eq:preq13} and
$g = G_f[m,n;t]$.
Combining both terms, the per-bin surrogate cost becomes
\begin{align}
\Phi_{m,n}(g)
&=
|1 - g H_f[m,n; t]|^2
\nonumber\\
&\quad+
(
\gamma
- \lambda_{\mathrm{ISAC}} S_f[m,n; t]\,
      |H_f[m,n; t]|^2
)
|g|^2.
\label{eq:perbin_cost}
\end{align}

Ignoring higher-order cross-bin couplings, the optimization
decouples across $(m,n)$ and the minimizer of
\eqref{eq:perbin_cost} satisfies
$\partial \Phi_{m,n}(g) / \partial g^{*} = 0$, which gives
\begin{equation}
G^{\mathrm{isac}}_f[m,n; t]
=
\frac{
  H_f^*[m,n; t]
}{
  |H_f[m,n; t]|^2(1 - \lambda_{\mathrm{ISAC}} S_f[m,n; t])
  + \gamma
}.
\label{eq:Gisac_final}
\end{equation}

In practice, $S_f[m,n;t]$ is obtained numerically:
we approximate the gradient magnitude
$|\partial J_{\mathrm{sense}}/\partial q_{m,n}|$
via small finite-difference perturbations of $\boldsymbol{\eta}$
and the induced squared changes in the effective AFDM kernel
$H_f[m,n;t]$, and then normalize $S_f[m,n;t]$ to lie in $[0,1]$
over the delay--Doppler grid.
To avoid noise enhancement when $|H_f[m,n;t]|^2$ is small
and $S_f[m,n;t]$ is large, we impose a small positive lower
bound on the denominator in \eqref{eq:Gisac_final}, on the
order of $\gamma$.
The time-domain pre-equalizer $G^{\mathrm{isac}}[\ell,n;t]$
is obtained by 2D inverse DFT of \eqref{eq:Gisac_final},
followed by the same per-frame transmit-power normalization
used for $G^{\mathrm{mmse}}[\ell,n;t]$.

\subsection{CSI variants and complexity}

At each time instant \(t\), Stage~II instantiates the AFDM-domain kernel \(H[\ell,n;t]\) from two CSI sources, both mapped via the reconstruction: (i) predicted CSI produced by Stage~I, and (ii) outdated CSI taken from the previous snapshot \((t-\Delta t)\). The instantaneous ground-truth parameters at \(t\) are mapped to reference kernels for error metrics and visualization in Section~VI but are not used to design pre-equalizers. All cases share the same grid size \((M,N)\), resolutions \((\Delta\tau,\Delta\nu)\), and the same AFDM phase-coupling structure, ensuring a consistent AFDM representation.

% Computationally, Stage~I is dominated by the GRU-based path predictor and the per-snapshot reconstruction of $K$ paths into an AFDM kernel.
% Stage~II uses only a small number of $M\times N$ 2D FFT/IFFT operations and $O(MN)$ pointwise operations per frame to evaluate either \eqref{eq:preq13} or \eqref{eq:Gisac_final}.

% The sensitivity map $S_f[m,n;t]$ and the corresponding FIM-based sensing cost are computed offline for representative channel snapshots and reused at run time. Although calculated offline, the sparsity of AFDM kernels suggests that a map obtained from representative snapshots can generalize reasonably well across local geometric variations within a pass. This reuse keeps the additional online overhead of the ISAC pre-equalizer compared with the classical MMSE case negligible, while low-complexity online FIM tracking or periodic sensitivity-map updates remain an interesting direction for future work.

Computationally, Stage I is dominated by GRU inference and reconstructing $K$ paths into an AFDM kernel. Stage II requires only a few $M\times N$ 2D FFT/IFFT operations and $O(MN)$ pointwise operations to evaluate either \eqref{eq:preq13} or \eqref{eq:Gisac_final}. The sensitivity map $S_f[m,n;t]$ (and the corresponding FIM-based sensing cost) is computed offline for representative snapshots and reused at run time, keeping the additional online overhead over the classical MMSE case negligible.

\section{Numerical Results and Discussion}

\subsection{Stage I: data-driven channel prediction}

We first evaluate the data-driven channel predictor on three disjoint 
LEO ISL passes (Channel~1, Channel~2, Channel~3). These passes were 
generated under different geometries and scattering conditions: Channel~2 
is dominated by a single strong cluster with relatively concentrated Doppler; 
Channel~3 exhibits stronger multi-cluster spreading and faster Doppler drift; 
Channel~1 sits in between and represents the most typical regime we observe 
in practice. Because Channel~1 is neither the easiest nor the hardest case, 
it is used as the default example in all subsequent visual figures.

Each ``channel'' in Table~\ref{tab:pred_accuracy} is not a single snapshot. 
Instead, it is an entire pass containing thousands of short-term frames. 
For every frame we reconstruct the delay--Doppler kernel from the network's 
predicted path parameters, compute the complex-valued normalized MSE (CNMSE), 
and then summarize that frame-wise error distribution by its mean, median 
(p50), and 90th percentile (p90). We also report the symmetric mean absolute 
percentage error of predicted path amplitudes (SMAPE\_amp) and the mean 
absolute phase error (MAE\_phase, in degrees). p50 and p90 summarize both central tendency and tail behavior of the frame-wise error distribution within each pass.
\begin{table}[htbp]
\centering
\footnotesize
\setlength{\tabcolsep}{4pt}
\caption{Prediction accuracy on three held-out LEO passes. Lower is better.}
\label{tab:pred_accuracy}
\begin{tabular}{lcccc}
\hline
Metric & Channel~1 & Channel~2 & Channel~3 & Mean \\
\hline
CNMSE mean (\%)      & 5.09  & 4.06  & 8.92  & 6.03 \\
CNMSE p50 (\%)       & 2.80  & 2.14  & 4.08  & 3.01 \\
CNMSE p90 (\%)       & 12.31 & 10.81 & 20.44 & 14.52 \\
SMAPE\_amp (\%)      & 2.37  & 1.43  & 2.55  & 2.12 \\
MAE\_phase (deg)     & 22.99 & 22.48 & 28.04 & 24.50 \\
\hline
\end{tabular}
\end{table}

\begin{figure*}[!t]
  \centering
  \includegraphics[width=\textwidth]{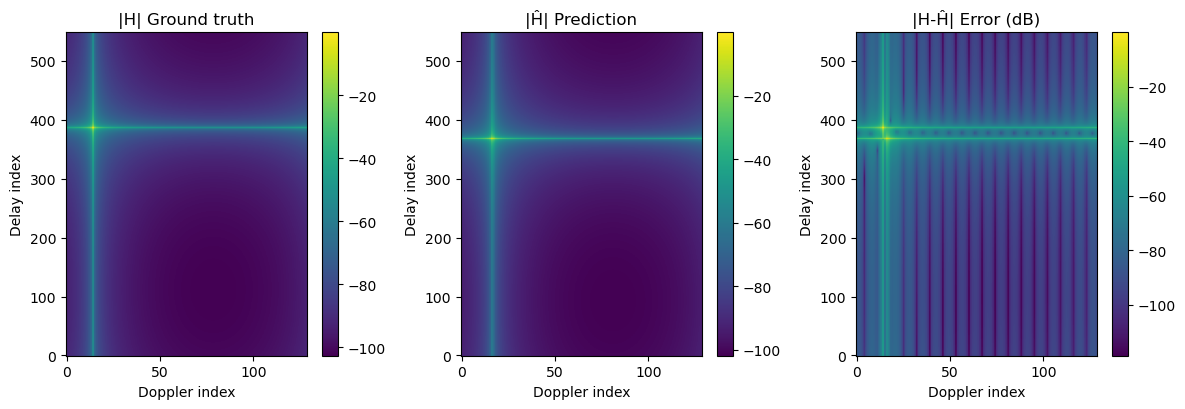}%
  \caption{Delay--Doppler magnitude and prediction error for one snapshot from Channel~1. Left: $|H|$; middle: $|\hat{H}|$; right: $|H-\hat{H}|$ in dB.}
  \label{fig:delay_doppler_errors}
\end{figure*}

Table~\ref{tab:pred_accuracy} shows that the predictor consistently achieves 
about 6\% CNMSE on average across the three passes, with median CNMSE near 
3\%. Even in the most challenging pass (Channel~3), which exhibits fast 
Doppler drift and richer multi-path, the 90th-percentile CNMSE stays on 
the order of 20\%, rather than diverging. The amplitude SMAPE is around 
2\%, and the mean absolute phase error is around 25 degrees. While a 
25-degree phase error may appear large in isolation, our downstream use 
case does not require sub-degree absolute carrier phase alignment. What 
matters is that the predictor recovers the dominant delay--Doppler support, 
cluster energies, and Doppler spreading trends. Visual comparisons of the 
true delay--Doppler kernel, the predicted kernel, and their difference in 
dB confirm that the main energy clusters are accurately reconstructed and 
that most residual error is concentrated in very weak side-lobes.

In Stage II we feed the predicted channel (rather than the ground truth) directly into a sensing-aware MMSE precoder, so accurate reconstruction of the dominant cluster and confinement of residual error to weak side-lobes are essential.

\subsubsection{Spatial distribution of prediction errors in the delay--Doppler domain}

While Table~\ref{tab:pred_accuracy} summarizes aggregate error statistics, it does not yet reveal where these errors occur in the physical delay--Doppler plane. To gain this insight, Fig.~\ref{fig:delay_doppler_errors} visualizes one representative snapshot from Channel~1, which we use as the default example throughout Stage~I. The left panel shows the magnitude of the true delay--Doppler kernel \(|H(\tau,\nu)|\) in dB, the middle panel shows the magnitude of the reconstructed kernel \(|\hat{H}(\tau,\nu)|\) obtained from the predicted path parameters, and the right panel shows the magnitude of the residual \(|H-\hat{H}|\) in dB. All three panels share the same color scale so that differences are directly comparable.

Comparing the first two panels, we see that the dominant energy cluster is preserved by the predictor. The peak location along both the delay and Doppler axes matches the ground truth, and the overall spread of the main lobe is closely reproduced. This is consistent with the low CNMSE\_mean and SMAPE\_amp values in Table~\ref{tab:pred_accuracy}, and indicates that the network has learned the underlying geometric structure of the strongest paths in the AFDM domain. The error map in the right panel further shows that large residuals are mostly confined to low-power regions away from the main cluster and to weak side-lobes, whereas the error in the vicinity of the dominant cluster is substantially smaller. The predictor therefore tends to make mistakes on very weak but numerous tail components, rather than on the few paths that dominate link performance.

This spatial pattern is crucial for the subsequent control stage. The ISAC pre-equalizer in Section~V relies on the amplitude and phase of the dominant delay--Doppler cluster to allocate gain and apply phase compensation. If the predicted main cluster were systematically shifted in delay or Doppler, the precoder would be steered toward the wrong region of the channel, potentially degrading both communication and sensing. Fig.~\ref{fig:delay_doppler_errors} instead shows that, for a representative LEO ISL pass, the location and envelope of the main cluster are well aligned between \(H\) and \(\hat{H}\), and that most residual error is confined to components that have limited impact on link reliability. This supports our choice in Stage~II to treat the predicted CSI as the only available channel information when constructing the pre-equalizer and CRLB-based sensing metric, rather than assuming access to an unrealistically clean or selectively filtered ground-truth channel.

\subsubsection{Comparison of network architectures and selection of a robust predictor}

\begin{figure}[htbp]
  \centering
  \includegraphics[width=\columnwidth]{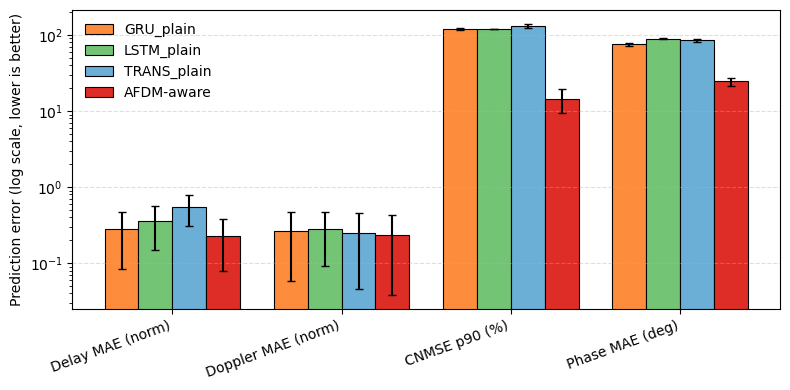}%
  \caption{Prediction errors of GRU\_plain, LSTM\_plain, TRANS\_plain, and AFDM-aware across four metrics (log scale).}

  \label{fig:arch_comparison}
\end{figure}

Before fixing the working point of the prediction module, we compare several network architectures on the same one-step prediction task. All models are trained with identical data splits and input–output formats: given the past 64 frames of AFDM-domain sparse path parameters, the network predicts the next frame of the $K$ strongest paths, including amplitude, phase, delay, and Doppler. To ensure a fair comparison, all baselines use the same normalization scheme in the AFDM parameter domain. In particular, the quantities Delay MAE (norm) and Doppler MAE (norm) in Fig.~\ref{fig:arch_comparison} denote the mean absolute error of the strongest-path delay and Doppler in this normalized coordinate system, rather than in physical units such as nanoseconds or hertz.

Fig.~\ref{fig:arch_comparison} summarizes the performance of three generic deep sequence models (GRU\_plain, LSTM\_plain, and TRANS\_plain) and the proposed AFDM-aware architecture with path tracking and physics-inspired regularization. Four key metrics are reported: strongest-path delay error, strongest-path Doppler error, the 90th-percentile complex NMSE (CNMSE p90), and strongest-path phase MAE. The horizontal axis enumerates these metrics, and the vertical axis uses a logarithmic scale. Each bar corresponds to the mean error over three unseen satellite ISL trajectories, while the error bars indicate the standard deviation across these channels, characterizing the cross-scenario robustness.

From Delay MAE (norm) and Doppler MAE (norm), we observe that the three generic sequence models achieve similar levels of normalized strongest-path prediction error, and the AFDM-aware architecture attains comparable or slightly lower errors on these metrics. This indicates that the proposed structure does not sacrifice short-term prediction accuracy. The differences become much more pronounced for the physically more critical metrics. For CNMSE p90, the generic models exhibit tail complex-channel errors on the order of $10^2$ percent, whereas the AFDM-aware predictor reduces the 90th-percentile error to the $10^1$ percent range, corresponding to nearly one order of magnitude improvement in the worst-case regime. A similar trend is observed for phase MAE: GRU, LSTM, and TRANS\_plain typically yield strongest-path phase errors around $80$--$90$ degrees, which is close to losing coherent phase information, while the AFDM-aware model reduces the phase MAE to roughly $20$--$30$ degrees, making subsequent coherent pre-equalization and beamforming still feasible.

The error bars further show that the AFDM-aware model achieves these lower mean errors without introducing larger variability across channels. On key metrics such as CNMSE p90 and phase MAE, its standard deviation is comparable to or smaller than that of the generic baselines. This indicates that the AFDM-aware architecture is not only superior on a single trajectory, but also more robust under different orbital geometries and scattering conditions. Therefore, we adopt the AFDM-aware predictor as the CSI source for all subsequent Stage II experiments.

\subsubsection{Path-level delay--Doppler consistency}

\begin{figure}[htbp]
  \centering
  \includegraphics[width=\columnwidth]{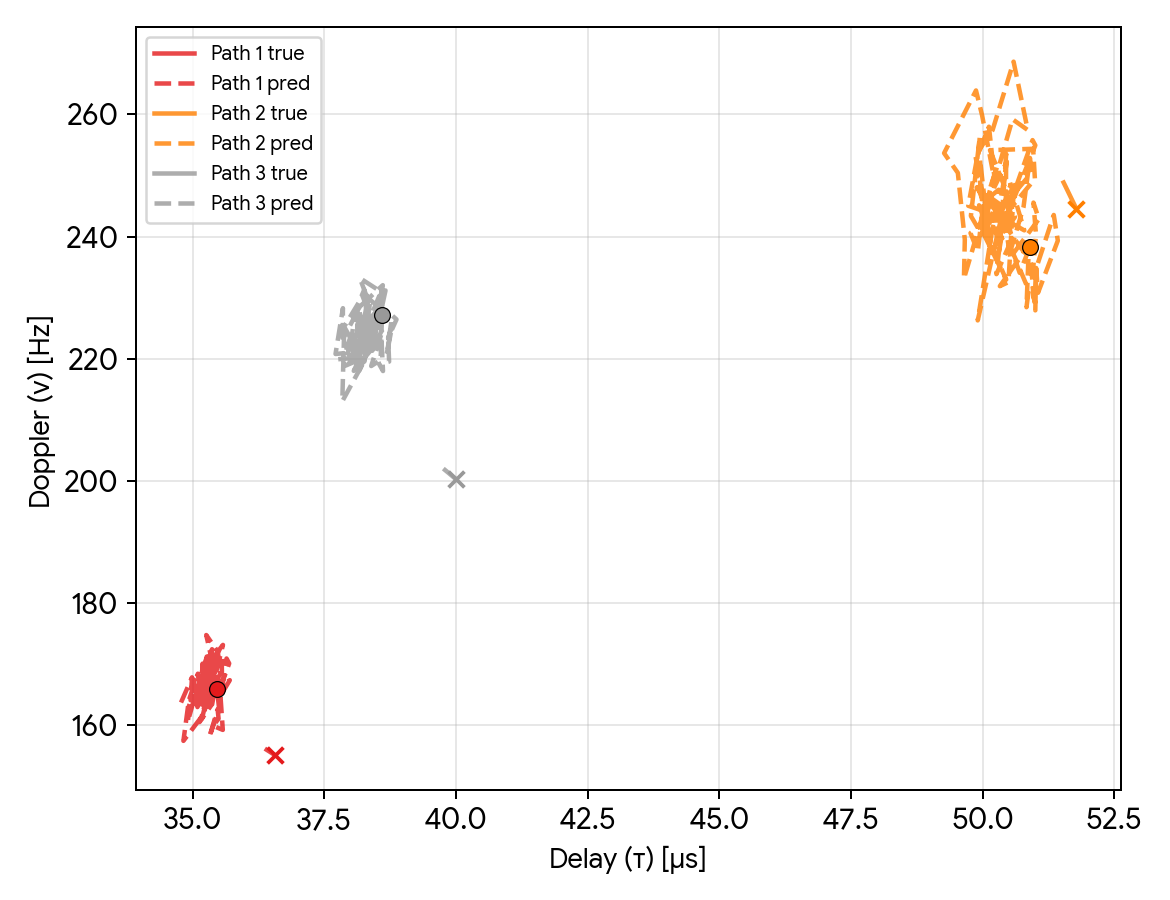}%
  \caption{Delay--Doppler trajectories of the three strongest paths in Channel~1. Solid curves and circles are true paths; dashed curves and crosses are predictions.}
  \label{fig:path_consistency}
\end{figure}

Table~\ref{tab:pred_accuracy} and Figs.~\ref{fig:delay_doppler_errors}--\ref{fig:arch_comparison} mainly evaluate prediction quality in terms of aggregate errors and energy distributions. To further verify that the model recovers physically meaningful multipath parameters, Fig.~\ref{fig:path_consistency} examines the consistency of the estimated delays and Dopplers at the path level. Specifically, we select the three strongest resolvable paths from Channel~1 and collect their estimates over time. The corresponding samples are plotted in the delay--Doppler plane: solid lines denote the true path trajectories, dashed lines denote the predicted trajectories, and solid circles and crosses mark the end points of the true and predicted trajectories at the final time step, respectively. The horizontal axis shows delay in microseconds, and the vertical axis shows Doppler in hertz.

Each true path forms a relatively tight cluster in the delay--Doppler plane, and the predicted trajectory closely envelopes this cluster. For all three paths, the center of the predicted cluster almost coincides with the true center; the delay bias is much smaller than the resolvable separation between paths, and the Doppler bias is well below the path's own drift range. The predictor does not simply reproduce the coarse power support, but also maintains consistent identity tracking of each dominant path in terms of delay and Doppler, without noticeable path swaps or large offsets. The temporal fluctuations of the dashed curves closely follow those of the solid curves, indicating that the model captures the slow drift of each path rather than producing a static biased estimate.

This path-level consistency is important for the subsequent ISAC pre-equalization stage. The Stage II precoder directly relies on per-path delay, Doppler, and phase parameters to allocate gain and apply coherent compensation. If the prediction were only shape-wise similar in terms of power but severely misaligned in the underlying parameters, any parameter-based control strategy would fail. Fig.~\ref{fig:path_consistency} shows that, for a representative LEO ISL trajectory, the predicted multipath parameters can be treated as a stable set of physically meaningful state variables that can be directly consumed by the control module, rather than as opaque high-dimensional deep features.

%\subsection{Summary of Stage~I channel prediction results}
\begin{figure*}[!t]
  \centering
  \includegraphics[width=0.9\textwidth]{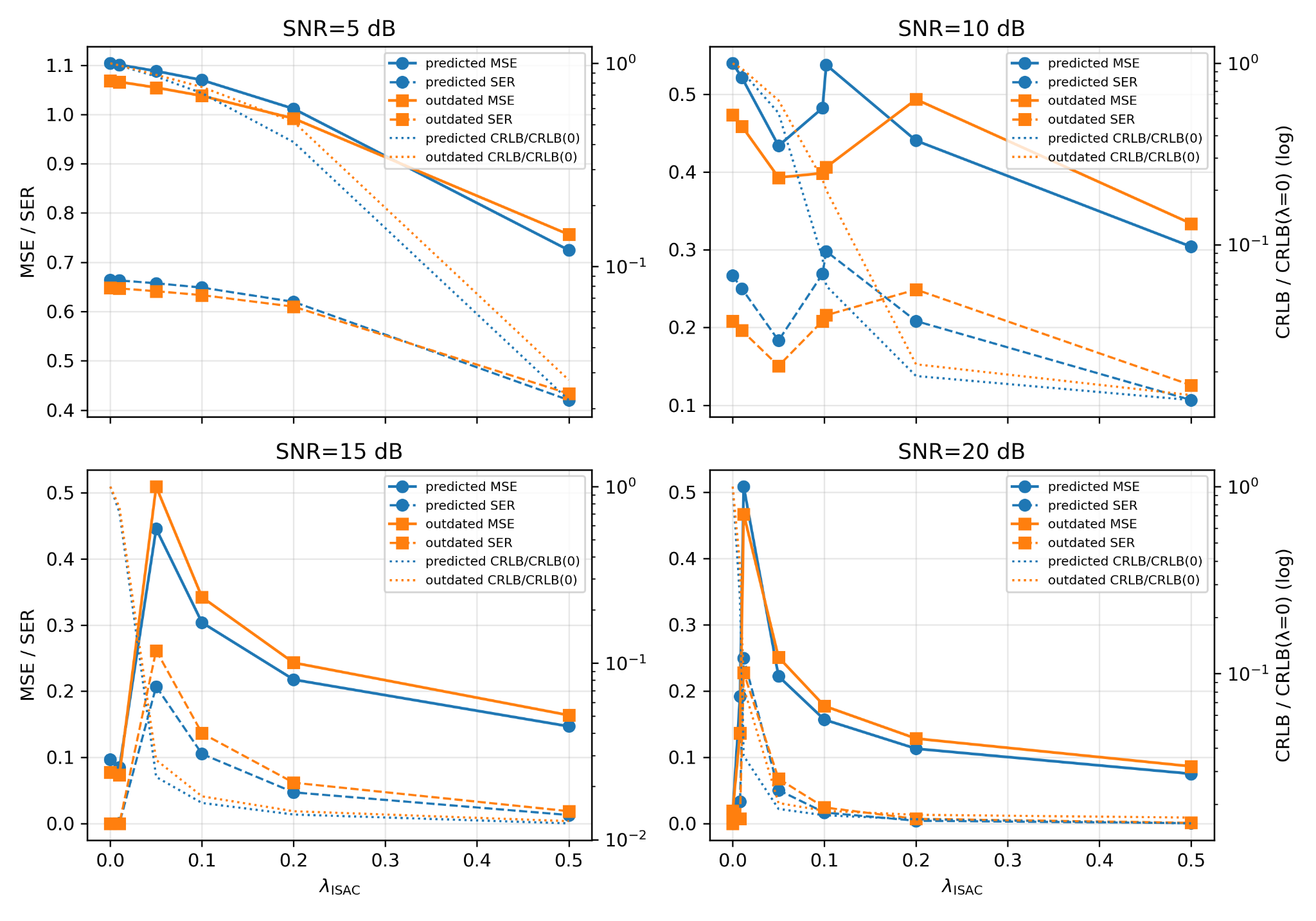}%
  \caption{Communication errors and normalized CRLB versus $\lambda_{\mathrm{ISAC}}$ for predicted and outdated CSI at several SNR values.}
  \vspace{-4mm}
  \label{fig:lambda_sweep}
\end{figure*}
%In summary, the Stage~I results demonstrate that the AFDM-domain path-parameter predictor provides both usable and interpretable CSI across multiple operating regimes. First, the statistics in Table~\ref{tab:pred_accuracy} indicate that, over three distinct LEO passes, the average complex NMSE is around 6\%, the median CNMSE is about 3\%, the amplitude SMAPE is on the order of 2\%, and the strongest-path phase MAE is around 20--30 degrees, keeping the overall error at a manageable level. Second, Fig.~\ref{fig:delay_doppler_errors} shows that most residual error is concentrated in weak side-lobe regions away from the dominant energy cluster, while the location and envelope of the main cluster in the delay--Doppler plane are well preserved, meaning that the structures that drive link performance are accurately recovered. Third, the architectural comparison in Fig.~\ref{fig:arch_comparison} reveals that, compared with generic GRU, LSTM, and Transformer sequence models, the AFDM-aware architecture with path tracking and physics-inspired regularization achieves an order-of-magnitude improvement in tail complex-channel error (CNMSE p90) and substantially lower phase MAE, without increasing variability across channels, thus providing a more robust working point. Finally, Fig.~\ref{fig:path_consistency} confirms at the path level that the predicted delays and Dopplers of the dominant paths closely follow their true trajectories, so that the network outputs can be regarded as physically meaningful state variables rather than opaque latent codes.

Based on these observations, all subsequent ISAC pre-equalization experiments use the AFDM-aware predictor as the only source of instantaneous CSI at time~$t$ for the proposed pre-equalizer; ground-truth path parameters and kernels are used only to construct the outdated-CSI baseline from $(t-\Delta t)$ and for evaluation. Consequently, the communication–sensing trade-offs reported in Stage II are obtained under the realistic assumption of data-driven predicted CSI, rather than under an optimistic measured-CSI benchmark. The three test channels span different levels of delay–Doppler sparsity and prediction difficulty, and the trade-off curves in Fig.~6 exhibit qualitatively similar behavior across them, which suggests that the proposed pre-equalizer is reasonably robust to moderate prediction errors; a more systematic analysis of error propagation is left to future work.

\subsection{Stage II: ISAC pre-equalizer under predicted CSI}

\subsubsection{\(\lambda_{\mathrm{ISAC}}\) sweep and communication–sensing tradeoff}

In Stage~II, the predicted CSI produced by Stage~I is used as the only
instantaneous channel information for the sensing-aware MMSE precoder
introduced in Section~V. The scalar weight \(\lambda_{\mathrm{ISAC}}\)
controls the tradeoff between pure communication performance and
sensing-oriented Fisher/CRLB metrics. For reference, Fig.~\ref{fig:lambda_sweep} also
includes a baseline where the precoder is designed from outdated CSI
(previous frame). Both cases use the same AFDM configuration, transmit
power, and noise power; \(\lambda_{\mathrm{ISAC}}\) only reshapes the
normalized effective channel \(H_{\mathrm{eff}}\) and does not change
the effective SNR by trivial power scaling. The sensing metric on the
right vertical axis is the normalized ratio
\(\mathrm{CRLB}(\lambda) / \mathrm{CRLB}(0)\) plotted on a logarithmic
scale, so that improvements at different SNR values can be compared.

At SNR \(= 5\)~dB, increasing \(\lambda_{\mathrm{ISAC}}\) from 0 to 0.5
reduces both the MSE and the SER monotonically while the normalized
CRLB decreases by nearly an order of magnitude. The curves based on
predicted and outdated CSI are almost parallel and numerically close.
Thus, in the low-SNR regime, introducing a sensing weight does not
force a tradeoff; instead, the same power budget can simultaneously
reduce detection error and estimation variance, yielding a favorable joint operating regime.

At moderate SNR values (10–15~dB), the curves take a more classical
tradeoff shape. For each SNR, when \(\lambda_{\mathrm{ISAC}}\) is
increased from 0 to roughly 0.05–0.1, the normalized CRLB already
drops by about one order of magnitude, whereas the MSE and SER remain
close to their \(\lambda_{\mathrm{ISAC}}=0\) baselines, with only
limited fluctuations. These small non-monotonic variations are mainly
caused by finite-sample effects and discrete modulation thresholds,
rather than by a change in the overall trend: there is a stable
operating interval where sensing observability can be significantly
enhanced without a noticeable loss in communication quality. In this
interval, the outdated-CSI baseline systematically exhibits slightly
higher MSE and SER than the predicted-CSI design, especially when
\(\lambda_{\mathrm{ISAC}}\) is larger, indicating that more up-to-date
CSI supports more reliable joint communication–sensing design under
the same weighting.

In the high-SNR regime at 20~dB, the sensing metric becomes even more
sensitive to \(\lambda_{\mathrm{ISAC}}\). Fig.~\ref{fig:lambda_sweep} shows that a small
deviation of \(\lambda_{\mathrm{ISAC}}\) away from zero already brings
\(\mathrm{CRLB}(\lambda) / \mathrm{CRLB}(0)\) down to the
\(10^{-2}\) range, while the MSE and SER enter a gently decreasing
plateau once \(\lambda_{\mathrm{ISAC}} \approx 0.05\). As in the
moderate-SNR case, the predicted-CSI curves are never worse than their
outdated-CSI counterparts over the entire sweep, and at larger
\(\lambda_{\mathrm{ISAC}}\) they maintain lower error levels for
comparable sensing gains.

Overall, Fig.~\ref{fig:lambda_sweep} indicates
that, under the given AFDM configuration and normalization strategy,
there exists a \(\lambda_{\mathrm{ISAC}}\) operating interval that
varies smoothly with SNR, within which the communication-side MSE and
SER deviate only mildly from the \(\lambda_{\mathrm{ISAC}}=0\) case
while the CRLB metric is substantially improved. Table~II summarizes
for each SNR a representative operating point
\(\lambda_{\mathrm{ISAC}}^\star\) chosen from this interval and
reports its statistics over three test ISL passes. As a result,
all Stage~II conclusions are drawn under a consistent predicted-CSI
assumption, and the comparison with outdated CSI shows that the gains
arise from the structured sensing-aware precoder rather than from
tuning a single SNR or a single snapshot in isolation.

\subsubsection{Statistical behavior of \(\lambda^\star\)}

Table~\ref{tab:lamstar_gain} summarizes, for each SNR, the distribution of the optimal sensing
weight \(\lambda^\star\) and the corresponding MSE-based synergy gain when the
precoder is designed from predicted CSI. For every pass and SNR, we scan
\(\lambda\) and define the gain \(g_{\mathrm{MSE}}\) as the relative MSE
improvement with respect to \(\lambda=0\). We report the median and interquartile range (IQR) of $\lambda^\star$ and $g_{\mathrm{MSE}}$ over three test passes ($n=3$); the associated SER and CRLB trends are visualized in Fig.~\ref{fig:lambda_sweep}; to keep the table compact we only retain the MSE-centric summary.

These results suggest an SNR-dependent operating policy. In the low-to-moderate SNR
regime (5–10~dB), the median \(\lambda^\star\) stays at relatively large values
(IQR collapsing to 0.50), and the median \(g_{\mathrm{MSE}}\) is about
0.34–0.42 with small cross-sample variation, indicating robust error reduction
from sensing-aware weighting. As SNR increases to 15~dB, the median
\(\lambda^\star\) drops to around 0.01 and the synergy gain decreases to the
\(7.7\times 10^{-2}\) range. At 20~dB, \(\lambda^\star\to 0\) and the MSE gain
vanishes, meaning the design naturally reverts to the pure communication MMSE
working point. Together with Fig.~\ref{fig:lambda_sweep}, these statistics show that a smoothly
varying \(\lambda\) schedule as a function of SNR provides a practical
selection rule: receiver satellites can look up or interpolate \(\lambda^\star\) from
SNR without per-link exhaustive search, while maintaining stable performance
across distinct ISL passes.

% ----------------- Table II -----------------
\begin{table}[htbp]
\centering
\footnotesize
\setlength{\tabcolsep}{4pt}
\caption{Optimal $\lambda^\star$ and MSE gain vs SNR (median [IQR], $n\!=\!3$)}
\label{tab:lamstar_gain}
\begin{tabular}{rcc}
\hline
SNR (dB) & $\lambda^\star$ median [IQR] & $g_{\mathrm{MSE}}$ median [IQR] \\
\hline
5  & 0.50 [0.50, 0.50] & 0.344 [0.34, 0.36] \\
10 & 0.50 [0.50, 0.50] & 0.421 [0.37, 0.43] \\
15 & 0.01 [0.01, 0.01] & 0.077 [0.05, 0.10] \\
20 & 0.00 [0.00, 0.00] & 0.000 [0.00, 0.00] \\
\hline
\end{tabular}
\end{table}

\subsubsection{Effective channel shaping driven by predicted CSI}
\begin{figure*}[htbp]
  \centering
  \includegraphics[width=\textwidth]{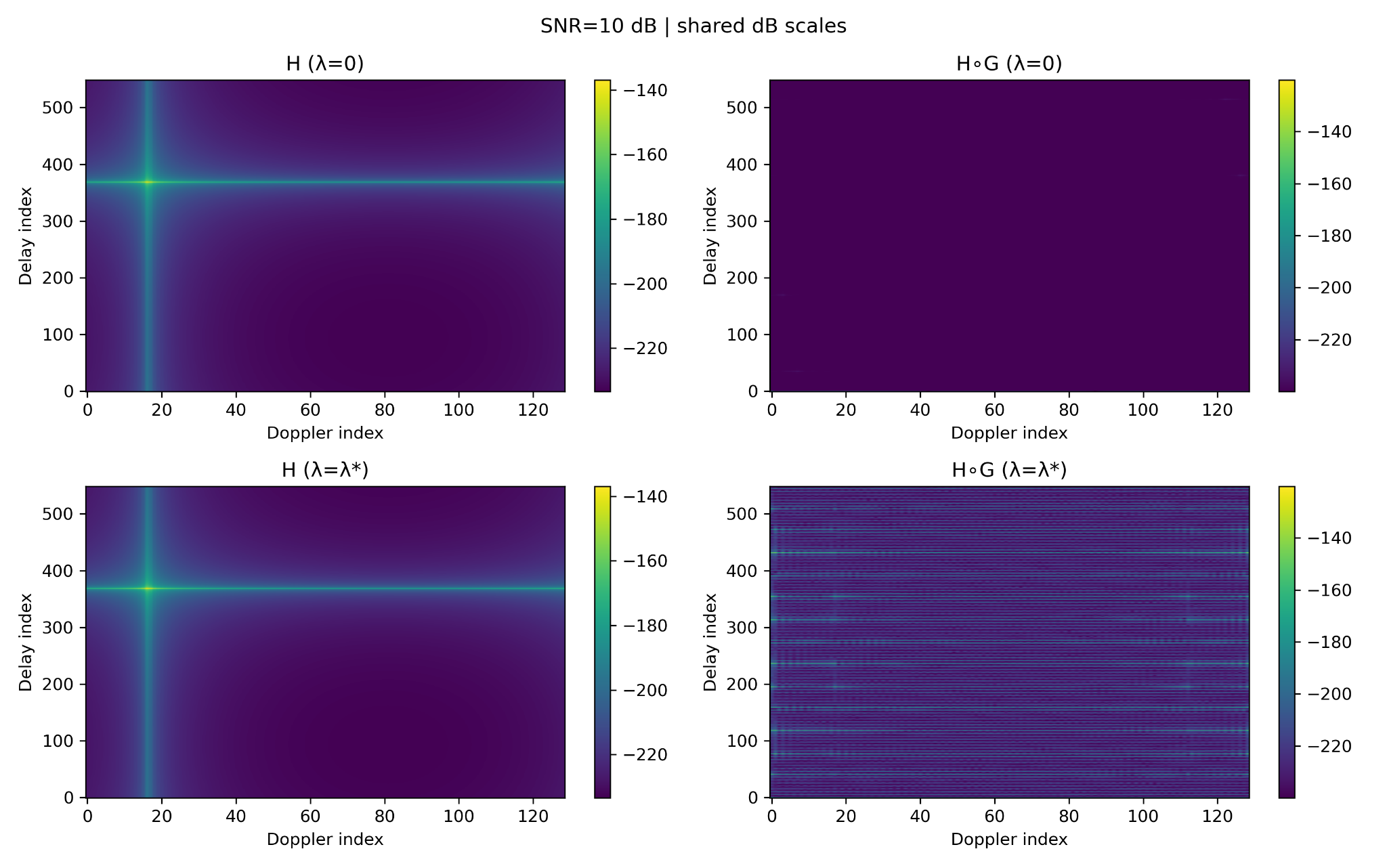}%
  \caption{Predicted channel $|H|$ and effective channel $|H \circledast G|$ at SNR $=10$ dB for $\lambda_{\mathrm{ISAC}}=0$ and $\lambda_{\mathrm{ISAC}}^\star$.}
  \label{fig:eff_channel_shaping}
\end{figure*}

Fig.~\ref{fig:eff_channel_shaping} illustrates, at SNR \(=10\) dB, how the sensing-aware precoder
shapes the effective channel in the delay–Doppler domain when it is
driven by predicted CSI. All four panels are generated from the same
predicted channel realization. The left column shows the magnitude of
the predicted kernel \(|H(\tau,\nu)|\); the right column shows the
magnitude of the effective kernel \(|H \circledast G(\tau,\nu)|\) seen after
precoding. The top row corresponds to \(\lambda_{\mathrm{ISAC}}=0\),
and the bottom row uses the operating point
\(\lambda_{\mathrm{ISAC}}^\star\) identified in Table~II for
10 dB. All panels share the same dB color scale, so that differences
reflect genuine reshaping of the kernel rather than trivial rescaling.

The two left panels are nearly identical: regardless of whether
\(\lambda_{\mathrm{ISAC}}=0\) or \(\lambda_{\mathrm{ISAC}}^\star\) is
used in the precoder, the predicted channel exhibits a dominant
delay–Doppler cluster at the same indices, with almost unchanged
envelope. This confirms that \(\lambda_{\mathrm{ISAC}}\) does not alter
the underlying CSI; it only affects the structure of the effective
kernel through the precoder.

In the communication-oriented case \(\lambda_{\mathrm{ISAC}}=0\)
(top row), the effective magnitude \(|H\circledast G|\) in the top-right
panel is strongly concentrated and largely pushed towards the bottom of
the shared color scale. Compared with the top-left \(|H|\), most
coefficients away from the dominant cluster are heavily suppressed,
while a small set of coefficients around the main support carries the
remaining energy. The resulting effective kernel resembles a
conventional MMSE equalizer that prioritizes flattening the strongest
components and suppressing residual multipath, thereby emphasizing
throughput and detection reliability over detailed structural
observability.

At the sensing-aware operating point
\(\lambda_{\mathrm{ISAC}}=\lambda_{\mathrm{ISAC}}^\star\) (bottom row),
the bottom-right panel shows a different pattern. The dominant
cluster remains clearly visible, but a much denser set of
delay–Doppler coefficients rises to intermediate amplitudes, forming a
structured, stripe-like footprint across the plane. Many coefficients
are lifted well above the noise floor compared with the
\(\lambda_{\mathrm{ISAC}}=0\) case, yet their magnitudes stay
significantly below the peak of the original \(|H|\), so the main
cluster still governs the overall gain. Because all four panels share
the same dB scale, this behavior can be directly interpreted as a
redistribution of energy across the delay–Doppler grid rather than as a
global power change.

This reshaping at 10 dB is consistent with the quantitative trends in
Fig.~\ref{fig:lambda_sweep} and Table~II. Around \(\lambda_{\mathrm{ISAC}}^\star\), the
precoder preserves the dominant cluster that supports reliable
communication, while allocating additional but controlled energy to
weaker cells that improve Fisher/CRLB-based observability. The MSE
penalty remains limited, whereas the sensing metric benefits from a
richer and more informative effective kernel. Fig.~\ref{fig:eff_channel_shaping} therefore
provides a structural view of how a predicted-CSI-driven ISAC precoder
interpolates between a pure communication equalizer
(\(\lambda_{\mathrm{ISAC}}=0\)) and a more distributed, sensing-friendly
footprint (\(\lambda_{\mathrm{ISAC}}^\star\)) on the same physical
channel.

\subsubsection{Ablation Study on the ISAC Precoder Design}

In this part, we assess the robustness and necessity of the proposed ISAC precoder by modifying only two key design degrees of freedom while keeping the predicted CSI, AFDM configuration, and simulation procedure unchanged:  
(i) the structure of the frequency-domain sensing weight map $S_f$;  
(ii) the normalization of the effective channel $H_{\mathrm{eff}}$.

We consider three ablation variants:
\begin{itemize}
  \item \textit{SfConst}: the derivative-based sensing map $S_f$ is replaced by an all-one constant, which corresponds to a spatially uniform regularization term.
  \item \textit{SfRandom}: $S_f$ is generated by multiplying the magnitude profile of the baseline derivative-based map $S_f$ with a fixed random pattern, without any physical structure constraint.
  \item \textit{noNorm}: the original $S_f$ is kept, but the normalization of $H_{\mathrm{eff}}$ is disabled in the communication simulations, so that $\lambda_{\mathrm{ISAC}}$ simultaneously changes the precoder shape and the effective SNR.
\end{itemize}

Table~\ref{tab:ablation_isac} summarizes the statistics of these three variants over the three test channels. For representative SNR values (5, 10, and 15 dB), we report (i) the median of the optimal tradeoff parameter $\lambda_{\mathrm{ISAC}}^\star$ and (ii) the median MSE synergy gain $g_{\mathrm{MSE}}$ with respect to $\lambda_{\mathrm{ISAC}}=0$. The fraction of channels that exhibit strictly positive gain at each SNR is discussed below. The proposed sensing-aware precoder serves as the baseline and its behavior across SNR is already reported in Table~\ref{tab:lamstar_gain}; Table~\ref{tab:ablation_isac} focuses on how the ablations deviate from this baseline.
\begin{table}[t]
  \centering
  \footnotesize
  \setlength{\tabcolsep}{4pt}
  \caption{Ablation study of the ISAC precoder design (median over three test channels).}
  \label{tab:ablation_isac}
  \begin{tabular}{c c c c}
    \hline
    Setting & SNR (dB) & $\lambda_{\mathrm{ISAC}}^\star$ (median) & MSE gain $g_{\mathrm{MSE}}$ (median) \\
    \hline
    SfConst  & 5  & 0.50  & 0.6640 \\
    SfConst  & 10 & 0.102 & 0.7977 \\
    SfConst  & 15 & 0.20  & 0.6012 \\
    SfRandom & 5  & 0.50  & 0.3546 \\
    SfRandom & 10 & 0.50  & 0.5870 \\
    SfRandom & 15 & 0.50  & 0.1846 \\
    noNorm   & 5  & 0.10  & 1.46$\times 10^{-4}$ \\
    noNorm   & 10 & 0.05  & 2.15$\times 10^{-4}$ \\
    noNorm   & 15 & 0.01  & 1.16$\times 10^{-4}$ \\
    \hline
  \end{tabular}
\end{table}

As shown in Table~\ref{tab:ablation_isac}, both \textit{SfConst} and \textit{SfRandom} can indeed provide non-negligible MSE improvements in the medium SNR range. For example, at SNR = 5--10 dB, their median $g_{\mathrm{MSE}}$ lies roughly between $0.35$ and $0.80$, and all three test channels benefit from a positive MSE gain at these SNRs. However, when compared with the smooth $\lambda_{\mathrm{ISAC}}^\star$--SNR trajectory of the proposed method in Table~II, these two ablations exhibit two undesirable behaviors. First, the optimal tradeoff parameter tends to saturate at the upper bound of the search interval (e.g., $\lambda_{\mathrm{ISAC}}^\star = 0.5$ for \textit{SfConst} and \textit{SfRandom} at 5 dB), indicating a preference for very strong, almost uniform regularization rather than a fine-grained communication–sensing tradeoff. Second, in the high-SNR regime, especially for \textit{SfRandom} at 20--25 dB, both $\lambda_{\mathrm{ISAC}}^\star$ and the median $g_{\mathrm{MSE}}$ collapse to zero, which suggests that purely random weighting fails to deliver stable and repeatable ISAC synergy when the SNR is high.

The \textit{noNorm} variant behaves almost like a degenerate case. Across all SNR values, its median MSE gain $g_{\mathrm{MSE}}$ in Table~\ref{tab:ablation_isac} remains on the order of $10^{-4}$, much smaller than the synergy gains reported for the proposed method in Table~\ref{tab:lamstar_gain}. In addition, in the 0--15 dB range, only two out of three channels exhibit strictly positive MSE improvement at 10 and 15 dB (all three channels still benefit at 5 dB). This indicates that, without normalizing $H_{\mathrm{eff}}$, most of the effect of $\lambda_{\mathrm{ISAC}}$ is absorbed by changes in the effective SNR, and the resulting tradeoff curves are neither robust nor easy to interpret.

Overall, Table~\ref{tab:ablation_isac} shows that simple constant or random choices of $S_f$ can yield some MSE reduction at specific SNRs, but they tend to push $\lambda_{\mathrm{ISAC}}^\star$ towards saturation or collapse and lose consistent synergy, especially at high SNR. Removing the $H_{\mathrm{eff}}$ normalization further destroys the ability to realize a controllable tradeoff via $\lambda_{\mathrm{ISAC}}$. Combined with the baseline results in Table~II, these ablation studies confirm that the gains in the second stage do not come from merely adding a regularization term; rather, they critically rely on the physically structured, derivative-based $S_f$ and a properly normalized simulation pipeline.

\section{Conclusion}

This paper has presented a two-stage AFDM-based ISAC framework for mobile LEO inter-satellite links operating under predicted CSI. Stage~I performs path-level prediction of a sparse set of dominant specular components and reconstructs the AFDM delay–Doppler kernel, which is used as the only instantaneous CSI at the transmitter. Stage~II augments conventional AFDM MMSE pre-equalization with a CRLB-inspired regularization, yielding an approximate closed-form pre-equalizer with a single parameter that balances communication MSE and sensing accuracy. Simulations over representative LEO ISL trajectories show that the proposed path-level predictor provides phase-consistent kernel reconstruction, and that the sensing-aware pre-equalizer driven by predicted CSI significantly improves sensing-oriented metrics over outdated-CSI baselines while keeping SER close to communication-oriented designs with modest extra complexity. Future work includes extensions to multi-antenna ISLs, more refined scattering models, and joint learning of the prediction and pre-equalization stages.

\ifCLASSOPTIONcaptionsoff
  \newpage
\fi

\bibliographystyle{IEEEtran}
\bibliography{IEEEabrv,Bibliography}

\vfill

\end{document}